\documentclass  {aip-cp}
\usepackage[numbers]{natbib}
\usepackage{rotating}
\usepackage{graphicx}
\usepackage{caption}
\usepackage{wrapfig}

\newcommand{\GERDA}       {\mbox{\textsc{Gerda}}}  
\begin{document}

\title{PEN as self-vetoing structural Material}



\author[aff1]{B. Majorovits\corref{cor1}}
\author[aff1]{S. Eck}
\author[aff1]{F. Fischer}
\author[aff1]{C. Gooch}
\author[aff2]{C. Hayward}
\author[aff1]{T. Kraetzschmar}
\author[aff1]{N.~van~der~Kolk}
\author[aff2]{D. Muenstermann}
\author[aff1]{O.~Schulz}
\author[aff1]{F. Simon}

\affil[aff1]{Max-Planck-Institut f\"ur Physik, F\"ohringer Ring 6, 80805 M\"unchen, Germany}
\affil[aff2]{Physics Department, Lancaster University, Lancaster, LA1 4YB, United Kingdom}
\corresp[cor1]{Corresponding author: bela.majorovits@mpp.mpg.de}

\maketitle

\begin{abstract}
Polyethylene Naphtalate (PEN) is a mechanically very favorable polymer. Earlier it was found
that thin foils made from PEN can have very high radio-purity compared to other commercially
available foils. In fact, PEN is already in use for low background signal transmission applications
(cables). Recently it has been realized that PEN also has favorable scintillating properties. In
combination, this makes PEN a very promising candidate as a self-vetoing structural material in
low background experiments. Components instrumented with light detectors could be built from
PEN. This includes detector holders, detector containments, signal transmission links, etc. The
current R\&D towards qualification of PEN as a self-vetoing low background structural material
is be presented. \end{abstract}

\maketitle

\vspace{0.5cm}
Experiments for the search of extremely rare events
such as neutrinoless double beta-decay or dark matter interactions require
extremely good control of the radio-purity 
of the materials they are made of.
Additionally, experiments can
also strongly benefit from instrumenting the detector
surroundings with materials that are 
scintillating and transparent for light, as recently
demonstrated by the GERDA experiment \cite{gerda_nature}.

\begin{wrapfigure}{l}{0.5\textwidth}
\captionsetup{justification=raggedright}
\captionsetup{labelformat=simple, labelsep=period}
\includegraphics[height=.3\textheight, angle=270]{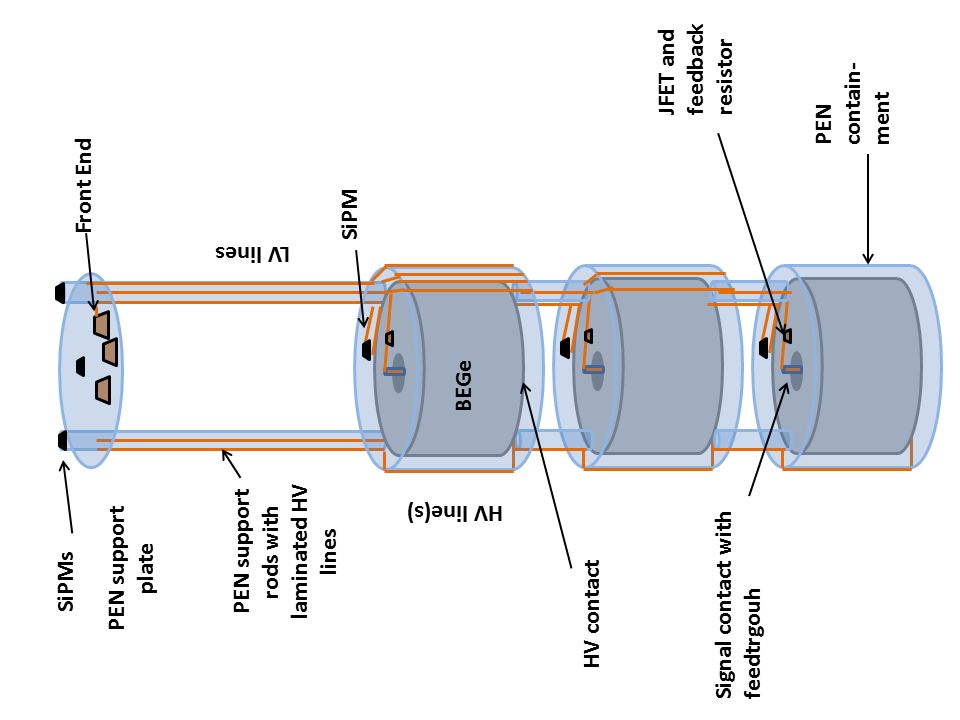}
  \caption{Sketch of a possible configuration of a germanium
  detector string in a future experiment.}
\label{fig:principle}
\end{wrapfigure}

Usually the infrastructural parts, such as
holders and containments are made of low
background copper or other 
materials known to be radio-pure.
These parts are, however, usually not transparent
for light, hence inactive from the viewpoint of background 
radiation detection.

Recently it has been reported that transparent 
Polyethylene Naphtalate (PEN), 
commonly used in industry, is 
efficiently scintillating with a 
peak emission at 425nm \cite{nakamura}. PEN has been
recognized earlier to be radio-pure and has already
been used in low background experiments as a cable 
substrate for signal transmision cables
\cite{cuore} and as a material for low background
HV capacitors \cite{capacitor}.

The mechanical properties of PEN at 
cryogenic temperatures (77 K) have been reported
to be rather favorable. The yield strength
has been shown to be $>$\,300 MPa, even for a 
degree of crystallinity of 44\% \cite{pen_mechanical}. 
This is a factor of $\approx$\,3 higher than for copper.
The Youngs modulus at 77\,K has been measured 
to be 13\,GPa, significantly 
higher than for other commonly used
polymers, such as polyethylene-teraphtalate. While this is considerably
less than for metals, it is still high enough
to allow for many standard mechanical applications.

The combination of its radio-purity, its favorable 
mechanical properties and the fact that
it is scintillating makes PEN a very interesting
material for applications in low background 
experiments. Generally, PEN could be used
as a structural material for detector holders
or containments. These components could be arranged 
with clever geometries, allowing to guide 
the scintillation light with high efficiency  to attached light detectors,
for example SiPMs. Like this the surrounding of
the detector could be practically fully self-vetoing. Additionally, the structural
components could also be used to detect radiation 
from  the surrounding and act as an additional 
veto system.

Ideally, PEN structural components could additionally
serve as the substrate for signal transmission lines
and HV leads. This could remove the usual challenge of 
having to find custom made low background cables and
largely reduce the complexity of proper cable routing 
and reduce microphonics due to lose cable strands.

A  simplistic sketch of a possible configuration of a string
containing three germanium detectors, encapsulated
in PEN instrumented with SiPMs is shown in Fig. \ref{fig:principle}.



PEN pellets of types TN-8050SC and TN-8065S and a 1\,mm thick sample
of PEN in the form of Scintirex$^{TM}$ were procured
from Tejin DuPont \cite{dupont}. 
A fraction of the pellets was sent for screening
measurements to the GeMPI HPGe screening facility at LNGS \cite{GeMPI}.
The results of these measurements are shown in the table below.

\begin{wraptable}{c}{0.4\textwidth}
\captionsetup{justification=raggedleft}
\captionsetup{labelformat=simple, labelsep=period}
\caption{\small \label{tab:radiopurity} Results of radio purity measurements obtained with a low background HPGe detector for two PEN samples. Uncertainties are given with approx. 68\% CL)} 
\hspace{0.25cm}
\begin{tabular}{l|cc}
       &  TN-8065S  & TN-8050SC \\
       & \multicolumn{2}{c}{[mBq/kg]}\\
\hline
Ra-228	&	$<$ 0.15 	  &    $<$ 0.15\\
Th-228	&	(0.23 $\pm$ 0.05) & $<$ 0.13\\
\hline
Ra-226		&(0.25 $\pm$ 0.05) & $<$ 0.11\\
Th-234	&	$<$ 11   & $<$ 15\\
Pa-234m&		$<$ 3.4  &  $<$3.0\\
\hline
U-235	&	$<$ 0.066 & $<$ 0.054\\
\hline
K-40:		&1600 $\pm$ 400 & 1000 $\pm$ 400\\
Cs-137	&	$<$ 0.057 & $<$ 0.064\\
\end{tabular}
\end{wraptable}


It is apparent that the samples are clean 
in terms of the uranium and thorium decay chains, while
there seems to be a contamination with $^{40}$K.
This might be due to the catalyst used during
PEN synthesis. 


The PEN pellets from Tejin Dupont were used to perform first test moldings 
at Fraunhofer Institute f\"ur Chemische Technologie (ICT) \cite{fraunhofer}.
In a first run, tiles with a size of 170x170x4\,mm$^3$ were molded in a
molding machine of type Ferromatic Milacron K110. 
There was no surface treatment of the mold, resulting in an undefined 
surface roughness.
Prior to molding, the pellets have been dried at 160$^{o}$C for 6 hours.
The mold temperature was kept at 30$^{o}$C, the 
temperature at the nozzle was between 280\,$^{o}$C and 300\,$^{o}$C.
Once the machine parameters were optimized,
air enclosure free and transparent tiles were produced.

\begin{figure}
\includegraphics[height=.25\textheight]{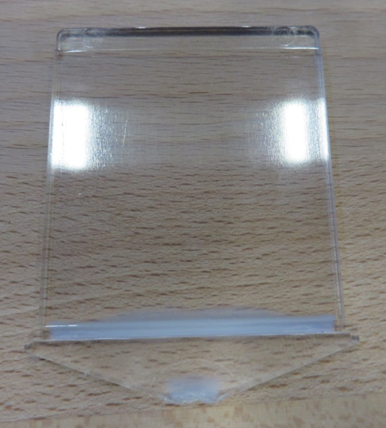}
\includegraphics[height=.25\textheight]{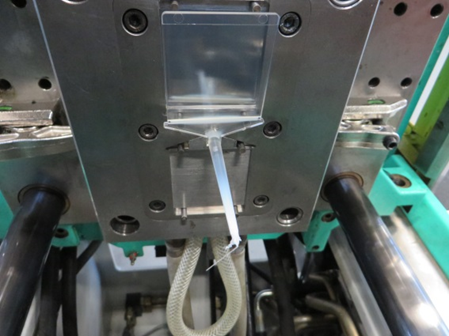}
  \caption{\label{fig:tiles} Tiles molded at Fraunhofer Institute ICT. Left: 170x170 mm$^2$ tile. Right: 55x50 mm$^2$ tile still inside the mold.}
\end{figure}

In another trial,  tiles
with size 50x55x3\,mm$^3$ were molded using a machine of type 
Arburg Allrounder 320C 600-250 using similar parameters as before.
Again, air enclosure free and transparent tiles were produced.
Both, pellets of types TN-8065S  and TN-8050SC were used.
Fig. \ref{fig:tiles} shows one tile of each geometry.

Independent molding tests have been performed at TU Dortmund \cite{stommel}. 
Transparent samples with geometry 30x30x3\,mm$^3$ could be molded succesfully.

For qualification of usage at cryogenic operational temperatures
the 170x170x4\,mm$^3$ tiles produced at Fraunhofer ICT
were immersed into liquid nitrogen $>$\,10 times.
Optical microscopy was used to determine differences
in the surface quality before and after thermal cycling.
Comparing pictures before and after cryogenic treatment 
no hints for material deterioration could be found.

For the characterization of scintillation properties, tiles with
standard geometry were produced from all samples. 
These have 30x30\,mm$^2$ surface
and a dimple with 10\,mm diameter and 1\,mm depth in the center 
on one side for
light collection with silicon photomultipliers (SiPMs), using the tile 
geometry developed for the CALICE analog hadron calorimeter
\cite{calice},\cite{yong}.

Samples molded at TU Dortmund were characterized at Lancaster University using a 
Agilent Cary Eclipse spectrograph \cite{connor_thesis}.
For emission measurements, the PEN samples were illuminated with a (360$\pm$2.5)\,nm light source,
the excitation light output was measured at 450\,nm.

\begin{figure}
\includegraphics[height=.225\textheight]{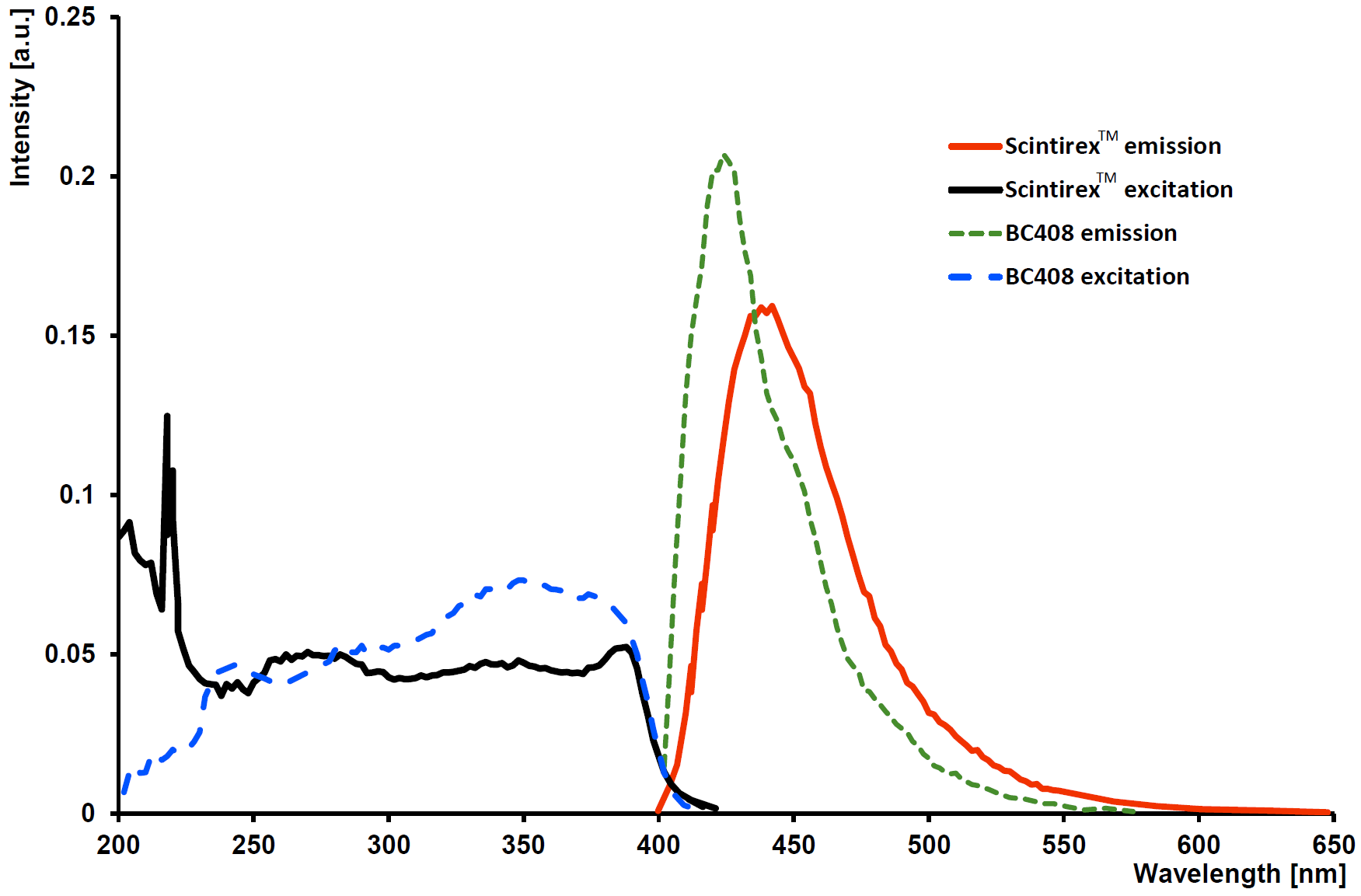}
\includegraphics[height=.225\textheight]{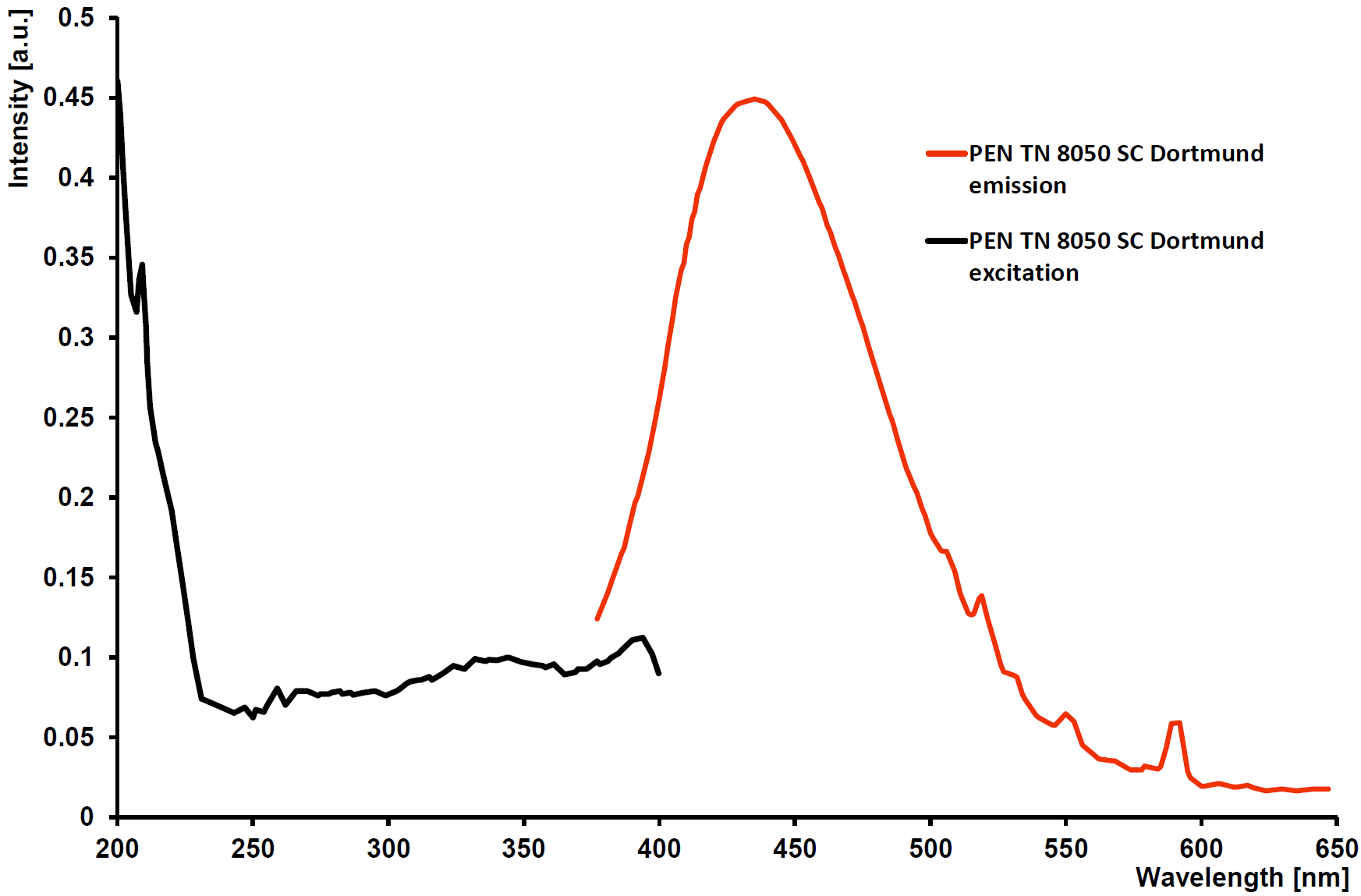}
  \caption{\label{fig:excitation_emission} Excitation and emission spectra of Scintirex$^{TM}$ (left) and TN-8050SC PEN molded at TU Dortmund (right). For comparison also the spectra for a BC408 sample are shown (left).}
\end{figure}

\begin{figure}[b!]
\includegraphics[height=.225\textheight]{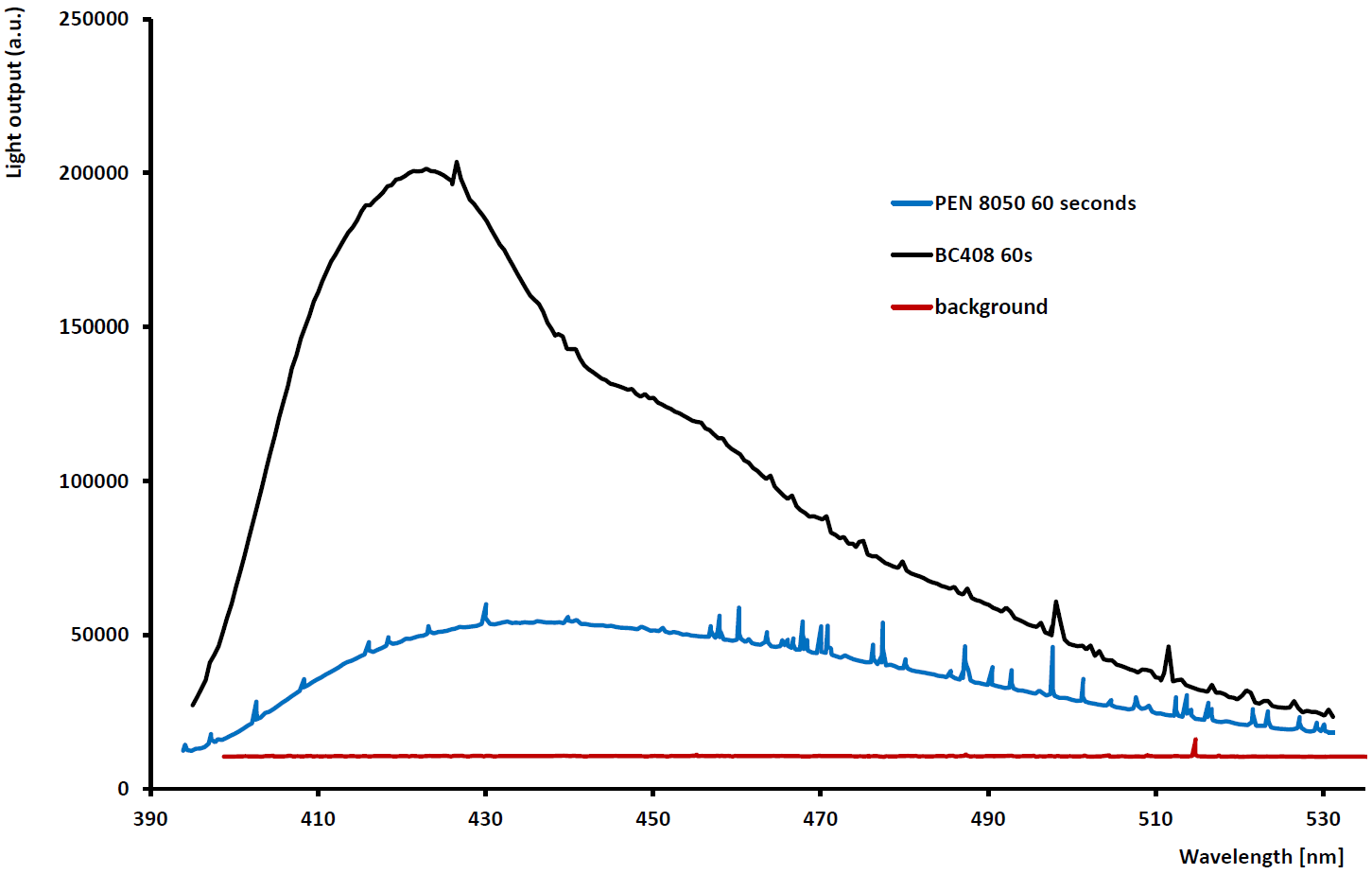}
\includegraphics[height=.225\textheight]{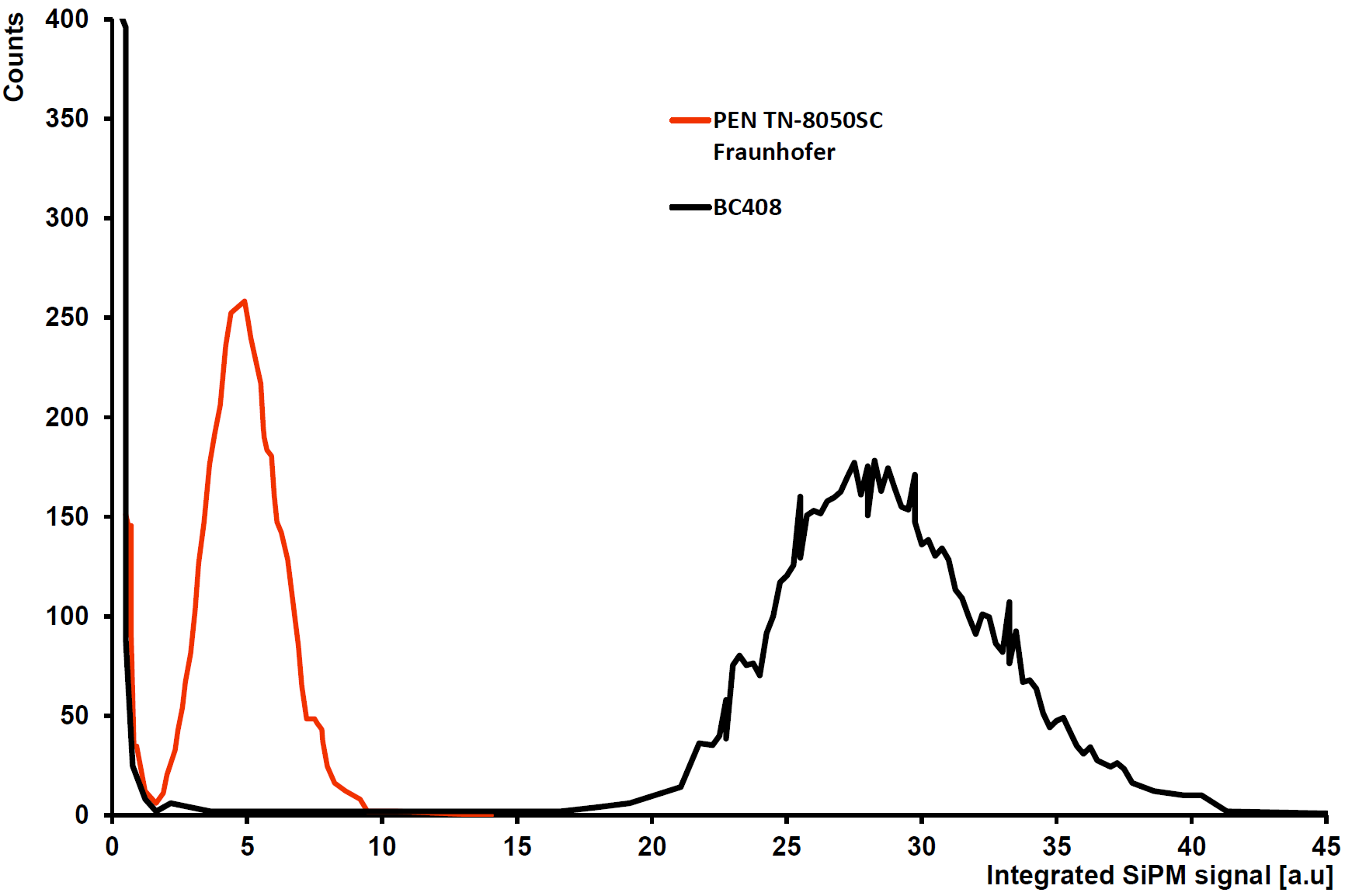}
  \caption{\label{fig:pen_bc408} Left: Comparison of scintillation light output as a function of frequency 
for two tiles made from BC408 standard scintillator and PEN TN-8050SC. Scintillation light was induced using a $^{106}$Ru source. A background measurement is also shown.   Right: Light output measurment of a PEN TN-8050SC tile and a BC 408 tile taken with SiPMs. A $^{90}$Sr source was used.}
\end{figure}

The Scintirex$^{TM}$ sample was investigated  using photo-luminescence at CERN. For comparison,
also a sample of standard scintillator from Saint Gobain BC408 \cite{bc408} with 3\,mm thickness was measured.
For emission measurements the Scintirex$^{TM}$ was illuminated with a (350$\pm$2.5)\,nm light source.
The excitation light output was measured at 420\,nm.
The obtained spectra are displayed in Fig. \ref{fig:excitation_emission}. 
It is clearly visible that all samples have an emission peak at wave lengths compatible with the detection peak efficiency of SiPMs.
The maximum light output was measured at 442\,nm and 435\,nm for the PEN TN-8050SC and the
Scintirex$^{TM}$ samples, respectively. This is both 
reasonably close to to the maximum emission wavelength of 425\,nm reported in \cite{nakamura}.
Remarkably, the overlap between excitation and emission spectra is as small as for BC408, showing
that PEN by itself acts as a wavelength shifter.
Also it is very interesting to note that the emission spectrum for both PEN samples (Scintirex$^{TM}$ and TN-8050SC) 
is increasing singnificantly below 230\,nm.

To prove that PEN can be used as an active detector component, a tile molded at Fraunhofer ICT from
PEN type TN-8050SC with geometry 30x30x3\,mm$^3$ with a dimple was exposed to a $^{106}$Ru $\beta$-source. 
Scintillation light produced inside the PEN tile was measured using an Andor Shamrock 303i spectrograph 
in connection with an Andor iDus 420 spectroscopy CCD \cite{andor}.
For comparison, the same measurement was repeated using a tile  of same geometry made from BC408. Also a blank measurement without tile and light source was performed.
Fig. \ref{fig:pen_bc408} (left) shows the resulting spectra. It is apparent that PEN emitts a significant 
amount of scintillation
light induced by $\beta$-particles, however, the light output of the sample was significantly less ($\approx$ 6 times) 
than for the BC408 tile.
 
Additionally a PEN tile molded at Fraunhofer ICT 
was exposed to a collimated $^{90}$Sr $\beta$ source in a dedicated scintillator test stand \cite{calice}.
The scintillation light was read out with a 3x3 mm$^2$ SiPM mounted inside the dimple on 
on the bottom of the 30x30x3 mm$^3$ tile. 
Only events that were triggered in a 5x5x5 mm$^3$ BC420 cube 
placed in the line of the $^{90}$Sr source below the PEN tile were recorded.
The same measurement was repeated with a tile of same geometry made from BC408.
The result of the measurement is displayed in Fig. \ref{fig:pen_bc408}.

In conclusion we state that first proof of principle tests seem to confirm
the potential of PEN as a self-vetoing structural material 
for low background applications: It could be shown that 
large enough forms can relatively easily be molded
from commercially available low background PEN pellets.
It was also confirmed that the samples molded from
these pellets are scintillating with a maximum
emission around 430\,nm.
Further investigations will need to be performed to qualify
PEN as a material to be used in future low background experiments.

\section{ACKNOWLEDGMENTS}
We would like to thank M. Stommel and M. Pohl from Technische Universit\"at Dortmund  as well as B. Beck and C. Anselment from Fraunhofer ICT for the work on molding the PEN tiles. Additionally we thank M. Laubenstein from Laboratorio Nazionale del Gran Sasso of the INFN for performing the PEN screening measurements. We thank E. Auffray from CERN for help with the photo-luminescence measurements and H. O'Keeffe for help with measurements at Lancaster University. This project was supported by the H2020 project AIDA-2020, GA no. 654168.



\bibliographystyle{aipproc}   




\end{document}